\documentclass[showpacs,amsmath,nofootinbib,amssymb,epsfig]{revtex4}

\usepackage{graphicx}
\usepackage{dcolumn}
\usepackage{bm}

\usepackage{color}
\usepackage{amssymb}

\begin{document}

\title{Geodesic Deviation Equation in $f(T)$ gravity}
\author{F. Darabi}
\email{f.darabi@azaruniv.ac.ir}\affiliation{Department of Physics, Azarbaijan Shahid Madani University , Tabriz, 53714-161 Iran}
\author{M. Mousavi}
\email{mousavi@azaruniv.ac.ir}\affiliation{Department of Physics, Azarbaijan Shahid Madani University , Tabriz, 53714-161 Iran}
\author{K. Atazadeh}
\email{atazadeh@azaruniv.ac.ir}\affiliation{Department of Physics, Azarbaijan Shahid Madani University , Tabriz, 53714-161 Iran}

\date{\today}

\begin{abstract}
In this work, we show that it is possible to  study the  notion of geodesic deviation equation in $f(T)$ gravity, in
spite of the fact that  in teleparallel gravity there is no notion of geodesics, and the torsion is responsible for the appearance of gravitational interaction. In this regard, we obtain the  GR equivalent equations for $f(T)$ gravity which are in the
modified  gravity form such as $f(R)$ gravity.
Then, we  obtain
the  GDE  within  the context of this modified gravity.
In this way, the obtained  geodesic deviation equation  will correspond to  the $f(T)$ gravity. Eventually, we extend the calculations to obtain the modification of Matting relation.
\end{abstract}
\vspace{1cm}
\maketitle

\section{Introduction}

The fundamental equation of Einstein geometrodynamics  and other metric theories of gravity is the Geodesic Deviation Equation(GDE)\cite{1}. It connects the spacetime curvature described by the Riemann tensor with a measurable physical quantity, namely the relative acceleration between two nearby test particles.  This equation describes the tendency of free falling particles to approach or recede from one another while moving under the influence of a spatially varying gravitational field. Actually the presence of this kind of tidal force will cause the trajectories to bend towards or away from each other which produces relative acceleration\cite{2}-\cite{3}. Moreover, the important Raychaudhuri equation and Mattig relation may be obtained by considering GDE equation for timelike and null congruences.

One  extend gravity theory beyond general relativity is
{\it teleparallel gravity} (TG). The birth of this gravity theory refers back
to 1928 \cite{6}. At that time Einstein was trying to redefine the unification of gravity and electromagnetism by introducing the notion of tetrad (vierbin) field together with the suggestion of absolute parallelism. In this theory the metric $g_{\mu\nu}$ is not the dynamical object, instead we have a set of tetrad fields $e_{a}(x^{\mu})$, and instead of the well-known torsionless Levi-Civita connection of GR theory, we work with a Weitzenb\"{o}k connection to introduce the covariant derivative \cite{Weitzenb}.
Furthermore, the role of curvature scalar in GR is played by torsion scalar $T$ in the teleparallel  gravity.

Although at the background and perturbation levels TG is completely equivalent to General
Relativity, $f(T)$ gravity has new structural and
phenomenological features.  Especially, at the cosmological
background it has various cosmological solutions which are consistent with the observational data \cite{Ferraro:2006jd,Ben09,Linder:2010py,Myrzakulov:2010vz,11,111,1111}. In addition, by taking spherical geometry one can
consider the spherical solutions for $f(T )$
gravity \cite{Wang,Mussa,MH,sari}. According to these features,  $f(T )$ gravity  is assumed as a viable theory both at cosmological and
at astrophysical aspects. Thus, the
cosmological and spherical solutions in $f(T)$ gravity lead to various viable
models that support cosmological observations \cite{111,Bas}
 along with Solar System tests \cite{1111} in which $f(T )$ must
be close to the linear form. In \cite{Wei,ata,Bas2} the authors
have studied the Noether Symmtery approach in the  Friedmann-Lema\^{\i}tre-Robertson-Walker (FLRW) geometry,
to construct some viable $f(T )$ functional forms \cite{capo2}. Regarding
this line of  progress in the context of $f(T)$ gravity, it seems there is still
room to study some motivating and interesting gravitational and cosmological aspects of $f(T)$ gravity which have not yet been studied.
 Following this idea, in this work we consider GDE in the context of $f(T)$ gravity.

The GDE has been studied  in $f(R)$ gravity theory \cite{fr1}, \cite{fr2},
so it is appealing to study the GDE in the context of $f(T)$ gravity, too. However, there are conceptual differences between GR and TG. In GR which
is fundamentally based on the weak equivalence principle, curvature is used
to geometrize the gravitational interaction and the spinless particles
follow the curvature of spacetime. In other words, the concept of force is
replaced by geometry and the particle trajectories   are determined by geodesics,
rather than the force equation. In TG, on the other hand, the torsion is responsible
for the appearance of   gravitational interaction as a real force. Hence,
there is no notion of geodesics in TG.  In spite of this conceptual difference,
one can show that the teleparallel description of the gravitational interaction
is completely equivalent to that of general relativity \cite{TR-GR}. Therefore,
it is possible to cast the force equation in TG into the form of a geodesic
equation in GR and obtain the corresponding GDE in TG.

In the present work,
we use another approach
 to obtain the
GDE in $f(T)$ gravity.
In this regard, first we use the method  introduced
in Refs \cite{barrow,barrow2,barrow3,reb} to obtain the GR equivalent of $f(T)$ gravity where the field equations are in the
modified gravity form, such as $f(R)$ gravity.
Then, we can benefit of the approach followed in \cite{fr1} to obtain
the GDE  within  the context of   $f(R)$ gravity.
In this way, we actually obtain the GDE  within  the context of   $f(T)$ gravity.\\

\section{Field equations in $f(T)$ gravity} \label{sec2}

Instead of using the torsionless Levi-Civita connection in General Relativity,
we use the curvatureless Weitzenb\"{o}ck connection in Teleparallelism
\cite{Weitzenb}, whose non-null torsion $T^\rho_{\verb| |\mu\nu}$ and contorsion
$K^{\rho}_{\verb| |\mu\nu}$ are defined respectively by
\begin{eqnarray}\label{eq1}
T^\rho_{\verb| |\mu\nu} \equiv \tilde{\Gamma}_{\nu\mu}^{\rho}-\tilde{\Gamma}_{\mu\nu}^{\rho}=e^\rho_A
\left( \partial_\mu e^A_\nu - \partial_\nu e^A_\mu \right),\,
\label{eq:2.2}
\end{eqnarray}
\begin{eqnarray}\label{eq2}
K_{\;\;\mu \nu }^{\rho} \equiv \tilde{\Gamma} _{\mu \nu }^{\rho}-\Gamma
{}_{\;\mu \nu }^{\rho}=\frac{1}{2}(T_{\mu }{}^{\rho }{}_{\nu }
+ T_{\nu}{}^{\rho }{}_{\mu }-T_{\;\;\mu \nu }^{\rho }),\,
\label{eq:2.3}
\end{eqnarray}
where  $\Gamma{}_{\;\;\mu \nu }^{\rho }$ is the
Levi-Civita connection.
Moreover, instead of the Ricci scalar $R$ for the Lagrangian density
in general relativity, the teleparallel Lagrangian density is described by the torsion scalar $T$ as follows
\begin{equation}\label{eq3}
T \equiv S_\rho^{\verb| |\mu\nu} T^\rho_{\verb| |\mu\nu}\, ,\,
\end{equation}
where
\begin{equation}\label{eq4}
S_\rho^{\verb| |\mu\nu} \equiv \frac{1}{2}
\left(K^{\mu\nu}_{\verb|  |\rho}+\delta^\mu_\rho \
T^{\alpha \nu}_{\verb|  |\alpha}-\delta^\nu_\rho \
T^{\alpha \mu}_{\verb|  |\alpha}\right)\,.
\end{equation}
%
The modified teleparallel action for $f(T)$ gravity
is given by~\cite{linder}
\begin{equation}\label{eq5}
{\cal S}=\frac{1}{2\kappa}\int
d^{4}x|e|{f(T)}+\int d^{4}x|e|{\cal L_{M}},
\end{equation}
where $|e|= \det \left(e^A_\mu \right)=\sqrt{-g}$, $8\pi G=\kappa$ and $c=1$.
Varying the action (\ref{eq1}) with respect to
the vierbein vector field $e_A^\mu$, we obtain the equation~\cite{rafael2}
\begin{eqnarray}\label{eq6}
&&\frac{1}{e} \partial_\mu \left( eS_A^{\verb| |\mu\nu} \right) f_{T}(T)
-e_A^\lambda T^\rho_{\verb| |\mu \lambda} S_\rho^{\verb| |\nu\mu}
f_{T}(T) + S_A^{\verb| |\mu\nu} \partial_\mu \left(T\right)f_{TT}(T)
+\frac{1}{4} e_A^\nu f = \kappa\Theta_A^\nu\,,
\end{eqnarray}
where a subscript $T$ denotes differentiation with respect to $T$ and $\Theta_A^\nu$ is the matter energy-momentum tensor, meanwhile all  indices on the manifold
 run over $0, 1, 2, 3$,
and $e_A^\mu$ form the tangent vector on the tangent space over which the
metric $\eta_{A B}$ is defined.

On the other hand, from the relation between the
Weitzenb\"{o}ck connection and the Levi-Civita connection
given by the equation  (\ref{eq:2.3}), one can write the Riemann tensor for
the Levi-Civita connection in the form

\begin{eqnarray}\label{tensorR}
R^\rho_{\;\;\mu\lambda\nu}\!\!\!\!\!\!&&=\partial_{\lambda}\Gamma{}_{\;\mu\nu}^{\rho}
-\partial_{\nu}\Gamma {}_{\;\mu\lambda}^{\rho}
+\Gamma {}_{\;\sigma\lambda}^{\rho}\Gamma {}_{\;\mu\nu}^{\sigma}
-\Gamma {}_{\;\sigma\nu}^{\rho}\Gamma {}_{\;\mu\lambda}^{\sigma}\\ \nonumber
&&=\nabla_\nu K^\rho_{\;\;\mu\lambda}-\nabla_\lambda K^\rho_{\;\;\mu\nu}
+K^\rho_{\;\;\sigma\nu}K^\sigma_{\;\;\mu\lambda}-K^\rho_{\;\;\sigma\lambda}K^\sigma_{\;\;\mu\nu}\;,
\end{eqnarray}
whose associated Ricci tensor can then be written as
\begin{equation}\label{eq8}
R_{\mu\nu}=\nabla_\nu K^\rho_{\;\;\mu\rho}-\nabla_\rho K^\rho_{\;\;\mu\nu}
+K^\rho_{\;\;\sigma\nu}K^\sigma_{\;\;\mu\rho}
-K^\rho_{\;\;\sigma\rho}K^\sigma_{\;\;\mu\nu}\;.
\end{equation}
Now, by using $K^\rho_{\;\;\mu\nu}$ given by the equation ~(\ref{eq2}) along with the
relations $K^{(\mu\nu)\sigma}=T^{\mu(\nu\sigma)}=S^{\mu(\nu\sigma)}=0$ and considering that $S^\mu_{\;\;\rho\mu}=
2K^\mu_{\;\;\;\rho\mu}=-2T^\mu_{\;\;\;\rho\mu}$ one has
~\cite{barrow,barrow2,barrow3,reb}
\begin{eqnarray}\label{eq9}
&&R_{\mu\nu}=-\nabla^\rho S_{\nu\rho\mu}-g_{\mu\nu}\nabla^\rho T^\sigma_{\;\;\;\rho\sigma}
-S^{\rho\sigma}_{\;\;\;\;\;\mu}K_{\sigma\rho\nu}\;, \nonumber \\
&&R=-T-2\nabla^\mu T^\nu_{\;\;\;\mu\nu}\;,
\end{eqnarray}
and thus obtain
\begin{equation}\label{eqdivs}
G_{\mu\nu}-\frac{1}{2}\,g_{\mu\nu}\,T
=-\nabla^\rho S_{\nu\rho\mu}-S^{\sigma\rho}_{\;\;\;\;\mu}K_{\rho\sigma\nu}\;,
\end{equation}
where $G_{\mu\nu}=R_{\mu\nu}-(1/2)\,g_{\mu\nu}\,R$ is the Einstein tensor.

Finally, by using the equation ~(\ref{eqdivs}), the field equations for
$f(T)$ gravity in terms of GR quantities, namely the equation ~(\ref{eq6}), can be rewritten in the following
form \cite{barrow,barrow2,barrow3,reb}

\begin{eqnarray}\label{eq11}
f_{T}G_{\mu\nu}+\frac{1}{2}\left(Tf_{T}-f(T)\right)g_{\mu\nu}+B_{\mu\nu}f_{TT}(T)=\kappa\Theta_{\mu\nu},
\end{eqnarray}
where $f_{T}=\frac{df(T)}{dT}$ , $f_{TT}(T)=\frac{df_{T}}{dT}$, $B_{\mu\nu}=S_{\nu\mu}\,^{\sigma}\nabla_{\sigma}T$ and $\Theta_{\mu\nu}$ is the matter energy-momentum tensor.
Now, this equation is in the form of field equation in modified gravity, such as $f(R)$ gravity.

\section{Geodesic Deviation Equation in GR}

Here we start with a little discussion about Geodesic Deviation Equation
in general relativity. The geometrical meaning of the Riemann tensor is best explained by examining the behavior of neighborhood geodesics. Imagine $C_{1}$ and $C_{2}$ are two adjacent geodesics with an affine parameter $\nu$ on 2-surface $S$ (see Fig.1).
The vector field $V^{\alpha}=\frac{dx^{\alpha}}{d\nu}$ is the normalized tangent vector of geodesic $C_{1}$, and $\eta^{\alpha}=\frac{dx^{\alpha}}{ds}$ is the deviation vector of these two adjacent geodesics. In total we describe these geodesics with $x^{\alpha}(\nu,s)$.\\

 Starting with $\pounds_{_{V}}\eta^{\alpha}=\pounds_{_{\eta}}V^{\alpha}$ ($[V,\eta]^{\alpha}=0$)
which leads to $\nabla_{_{V}}\nabla_{_{V}}\eta^{\alpha}=\nabla_{_{V}}\nabla_{\eta}V^{\alpha}$
and using $\nabla_{_{X}}\nabla_{_{Y}}Z^{\alpha}-\nabla_{_{Y}}\nabla_{X}Z^{\alpha}-\nabla_{_{[X,Y]}}Z^{\alpha}=
R^{\alpha}\,_{\beta\gamma\delta}Z^{\beta}X^{\gamma}Y^{\delta}$ in which $Y^{\alpha}=\eta^{\alpha}$ and $X^{\alpha}=Z^{\alpha}=V^{\alpha}$, we can obtain the GDE as follows \cite{3}
\begin{equation}\label{eq12}
\frac{D^{2}\eta^{\alpha}}{D\nu^{2}}=-R^{\alpha}\,_{\beta\gamma\delta}V^{\beta}\eta^{\gamma}V^{\delta},
\end{equation}

As an introduction, here we review briefly the results of finding GDE in GR.
We take  the energy momentum tensor in the
form of a perfect fluid \begin{equation}\label{eq14}
\Theta_{\mu\nu}=(\rho+p)u_{\alpha}u_{\beta}+pg_{\alpha\beta},
\end{equation}
where $\rho$ is the energy density and  $p$ is the pressure. The trace of energy momentum tensor is

\begin{equation}\label{eq15}
\Theta=3p-\rho.
\end{equation}
Now, we are supposed to calculate $R$ and $R_{\mu\nu}$ by looking at the standard form of the Einstein field equations in GR (with cosmological constant)

\begin{equation}\label{eq16}
R_{\mu\nu}-\frac{1}{2}Rg_{\mu\nu}+\Lambda g_{\mu\nu}=\kappa\Theta_{\mu\nu}.
\end{equation}
   The  Ricci scalar and Ricci tensor are obtained as follows

\begin{equation}\label{eq17}
R=\kappa (\rho-3p)+4\Lambda,
\end{equation}

\begin{equation}\label{eq18}
R_{\mu\nu}=\kappa (\rho+p)u_{\alpha}u_{\beta}+\frac{1}{2}[\kappa(\rho-p)+2\Lambda]g_{\mu\nu}.
\end{equation}
Considering these expressions and using the following equation\cite{3}

\begin{eqnarray}\label{eq19}
R_{\alpha\beta\gamma\delta}&=&\frac{1}{2}(g_{\alpha\gamma}R_{\delta\beta}-
g_{\alpha\delta}R_{\gamma\beta}+g_{\beta\delta}R_{\gamma\alpha}-g_{\beta\gamma}R_{\delta\alpha})- \frac{R}{6}(g_{\alpha\gamma}g_{\delta\beta}-
g_{\alpha\delta}g_{\gamma\beta})+C_{\alpha\beta\gamma\delta},
\end{eqnarray}
the right hand side of GDE is

\begin{eqnarray}\label{eq20}
R^{\alpha}\,_{\beta\gamma\delta}V^{\beta}\eta^{\gamma}V^{\delta}=\left[\frac{1}{3}(\kappa \rho+\Lambda)\epsilon +\frac{1}{2}\kappa (\rho+p)E^{2}\right]\eta^{\alpha},
\end{eqnarray}
where $\epsilon=V^{\alpha}V_{\alpha}$ and $E=-V_{\alpha}u_{\alpha}$. This equation is known as {\it Pirani equation} \cite{2}.
It is worth mentioning that one may derive the Pirani equation for every metric for which the Weyl tensor vanishes.
A great deal of important results from this equation have been obtained including some solutions for spacelike, timelike and null congruences  \cite{ellis}.

\begin{figure}[ht]
  \centering
  \includegraphics[width=4cm]{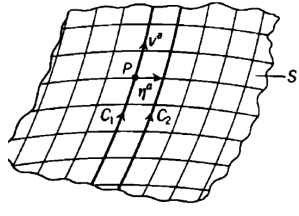}~
     \caption{Geodesic Deviation..}
  \label{fig:WEC1a}
\end{figure}

\section{Geodesics Deviation Equation in    $f(T)$ Gravity}

In this section, we shall obtain the GDE in GR equivalent theory of $f(T)$ gravity.
Before going through the details of calculations we extract $R$  by taking the trace of  (\ref{eq11})    which results in
\begin{eqnarray}\label{eq21}
R=\frac{1}{f_{T}}\left[2Tf(T)-2f(T)+f_{TT}S^{\mu}\,_{\mu\rho}\nabla^{\rho}T-\kappa\Theta\right].
\end{eqnarray}
Inserting this Ricci scalar into (\ref{eq11})
we obtain\begin{eqnarray}\label{eq22}
R_{\mu\nu}&=&\frac{1}{f_{T}}\Big[\frac{1}{2}g_{\mu\nu}(Tf_{T}-f+f_{TT}S^{\mu}\,_{\mu\rho}\nabla^{\rho}T-\kappa\Theta)-
f_{TT}S_{\nu\mu\rho}\nabla^{\rho}T+\kappa\Theta_{\mu\nu}\Big].
\end{eqnarray}
By applying the equation  (\ref{eq19}) and considering the zero value of the Weyl tensor $C_{\alpha\beta\gamma\delta}$ we will find.

\begin{widetext}
\begin{eqnarray}\label{eq23}
R_{\alpha\beta\gamma\delta}&=&\frac{1}{2f_{T}}\Big[\kappa(g_{\alpha\gamma}\Theta_{\delta\beta}-g_{\alpha\delta}\Theta_{\gamma\beta}+
g_{\beta\delta}\Theta_{\gamma\alpha}-g_{\beta\gamma}\Theta_{\delta\alpha})+(-f-\kappa \Theta+Tf_{T}+S^{\mu}\,_{\mu\rho}\nabla^{\rho}T f_{TT})(g_{\alpha\gamma} g_{\delta\beta}-g_{\alpha\delta}g_{\gamma\beta})\\\nonumber&&+(g_{\alpha\gamma}D_{\delta\beta}-
g_{\alpha\delta}D_{\gamma\beta}+g_{\beta\delta}D_{\gamma\alpha}-g_{\beta\gamma}D_{\delta\alpha})f_{T}\Big]-
\frac{1}{6f_{T}}\Big[-2f+2Tf_{T}+f_{TT}S^{\mu}_{\mu\rho}\nabla^{\rho}T-\kappa\Theta\Big](g_{\alpha\gamma}
g_{\delta\beta}-g_{\alpha\delta}g_{\gamma\beta}),
\end{eqnarray}
\end{widetext}
where
\begin{equation}\label{eq24}
D_{\mu\nu}=-S_{\nu\mu\rho} \nabla^{\rho}T \partial_{_{T}}.
\end{equation}
After rising $\alpha$ index in the Riemann tensor and contracting with $V^{\beta}\eta^{\gamma}V^{\delta}$ we will have

\begin{widetext}
\begin{eqnarray}\label{eq25}
R^{\alpha}\,_{\beta\gamma\delta}V^{\beta}\eta^{\gamma}V^{\delta}&=& \\ \nonumber
&&\frac{1}{2f_{T}}\Big[\kappa(\delta^{\alpha}_{\gamma}\Theta_{\delta\beta}-
\delta^{\alpha}_{\delta}\Theta_{\gamma\beta}
+g_{\beta\delta}\Theta^{\alpha}_{\gamma}-g_{\beta\gamma}\Theta^{\alpha}_{\delta})+(-f-\kappa\Theta+Tf_{T}+S^{\mu}\,_{\mu\rho}\nabla^{\rho}T f_{TT})(\delta^{\alpha}_{\gamma}g_{\delta\beta}-\delta^{\alpha}_{\delta}g_{\gamma\beta})\\\nonumber&&+
(\delta^{\alpha}_{\gamma}D_{\delta\beta}-\delta^{\alpha}_{\delta}D_{\gamma\beta}+
g_{\beta\delta}D^{\alpha}_{\gamma}-g_{\beta\gamma}D^{\alpha}_{\delta})f_{T}\Big]V^{\beta}\eta^{\gamma}V^{\delta}-
\frac{1}{6f_{T}}\Big[-2f+2Tf_{T}+f_{TT}S^{\mu}\,_{\mu\rho}\nabla^{\rho}T-\kappa\Theta\Big]\times\\\nonumber&&(g_{\alpha\gamma}g_{\delta\beta}-
g_{\alpha\delta}g_{\gamma\beta})V^{\beta}\eta^{\gamma}V^{\delta}.
\end{eqnarray}
\end{widetext}
Note that Eqs.(\ref{eq23}) and (\ref{eq25}) hold only if the Weyl tensor vanishes. By considering the equation  (\ref{eq14}) in the equation  (\ref{eq23}) we find
\begin{widetext}
\begin{eqnarray}\label{eq26}
R^{\alpha}\,_{\beta\gamma\delta}&=&\frac{1}{2f_{T}}\Big[\kappa(\rho+p)(g_{\alpha\gamma}u_{\delta}u_{\beta}-g_{\alpha\delta}u_{\gamma}u_{\beta}+
g_{\beta\delta}u_{\gamma}u_{\alpha}-g_{\beta\gamma}u_{\delta}u_{\alpha})+\\\nonumber
&&(g_{\alpha\gamma}g_{\delta\beta}-g_{\alpha\delta}g_{\gamma\beta})\left(\frac{2\kappa \rho}{3}+Tf_{T}/3 +
\frac{2}{3}S^{\mu}\,_{\mu\rho}\nabla^{\rho}T f_{TT}-\frac{f}{3}\right)
+(g_{\alpha\gamma}D_{\delta\beta}-g_{\alpha\delta}D_{\gamma\beta}+
g_{\beta\delta}D_{\gamma\alpha}-g_{\beta\gamma}D_{\delta\alpha})f_{T}\Big].
\end{eqnarray}
\end{widetext}
Under the condition of vector field normalization, we have $V^{\alpha}V_{\alpha}=\epsilon$ and
\begin{eqnarray}\label{eq27}
R_{\alpha\beta\gamma\delta}V^{\beta}V^{\delta}&=&\frac{1}{2f_{T}}[\kappa(\rho+p)(g_{\alpha\gamma}
(u_{\beta}V^{\beta})^{2}-2(u_{\beta}V^{\beta})V_{(\alpha}u_{\gamma)}+\epsilon u_{\alpha}u_{\gamma})+\\\nonumber&&\left(\frac{2\kappa \rho}{3}+\frac{Tf_{T}}{3}+\frac{2}{3}S^{\mu}\,_{\mu\rho}\nabla^{\rho}T f_{TT}-\frac{f}{3}\right)(\epsilon g_{\alpha\gamma}-V_{\alpha}V_{\gamma}) \\ \nonumber&& +[(g_{\alpha\gamma}D_{\delta\beta}-g_{\alpha\delta}D_{\gamma\beta}
+g_{\beta\delta}D_{\gamma\alpha}-g_{\beta\gamma}D_{\delta\alpha})f_{T}]V^{\beta}V^{\delta}].
\end{eqnarray}
Again we rise the first index then contract with $\eta^{\gamma}$
\begin{widetext}
\begin{eqnarray}\label{eq28}
R^{\alpha}\,_{\beta\gamma\delta}V^{\beta}\eta^{\gamma}V^{\delta}&=&\frac{1}{2f_{T}}\Big[\kappa(\rho+p)((u_{\beta}V^{\beta})^{2}\eta^{\alpha}-
(u_{\beta}V^{\beta})V^{\alpha}(u_{\gamma}\eta^{\gamma})-(u_{\beta}V^{\beta})u^{\alpha}(V_{\gamma}\eta^{\gamma})+\epsilon u_{\alpha}u_{\gamma}\eta^{\gamma})+\\\nonumber&&\left(\frac{2\kappa \rho}{3}+\frac{Tf_{T}}{3}+\frac{2}{3}S^{\mu}\,\,_{\mu\rho}\nabla^{\rho}T f_{TT}-\frac{f}{3}\right)(\epsilon \eta^{\alpha}-V_{\alpha}(V_{\gamma}\eta^{\gamma}))\\\nonumber&& +[(\delta^{\alpha}_{\gamma}D_{\delta\beta}-\delta^{\alpha}_{\delta}D_{\gamma\beta}+g_{\beta\delta}D^{\alpha}_{\gamma}-
g_{\beta\gamma}D^{\alpha}_{\delta})f_{T}]V^{\beta}V^{\delta}\eta^{\gamma}\Big].
\end{eqnarray}
\end{widetext}
By considering $E=-V_{\alpha}u^{\alpha}$ and $\eta_{\alpha}u^{\alpha}=\eta_{\alpha}V^{\alpha}=0$, the equation  (\ref{eq28}) converts into
\begin{widetext}
\begin{eqnarray}\label{eq29}
R^{\alpha}\,_{\beta\gamma\delta}V^{\beta}\eta^{\gamma}V^{\delta}&=&\frac{1}{2f_{T}}\Big[\kappa(\rho+p)E^{2}+\epsilon\Big(\frac{2\kappa \rho}{3}+\frac{Tf_{T}}{3}+\frac{2}{3}S^{\mu}\,_{\mu\rho}\nabla^{\rho}T f_{TT}-\frac{f}{3}\Big)\Big]\eta^{\alpha}\\\nonumber&& +\frac{1}{2f_{T}}\Big[(\delta^{\alpha}_{\gamma}D_{\delta\beta}-\delta^{\alpha}_{\delta}D_{\gamma\beta}+g_{\beta\delta}D^{\alpha}_{\gamma}-
g_{\beta\gamma}D^{\alpha}_{\delta})f_{T}]V^{\beta}V^{\delta}\Big]\eta^{\gamma}.
\end{eqnarray}
\end{widetext}

It should be mentioned that all the results obtained in this section do not require the FRW ansatz, rather are valid as long as the Weyl tensor vanishes.
In the next section, we will use these results to obtain the GDE in GR equivalent of $f(T)$ gravity model for FLRW metric whose Weyl tensor is zero. Obviously, our final result will be acceptable provided that the GR limit of this model be checked.

\subsection{  GR equivalent method with FLRW background}
We take the standard model line element (FLRW universe) including $a(t)$ and $k$, respectively  as scale factor and spatial curvature of the universe, as
 \begin{equation}\label{eq13}
 ds^{2}=-dt^{2}+a^{2}(t)\left[\frac{dr^{2}}{1-k r^{2}}+r^{2}d\theta^{2}+r^{2}sin^{2}\theta d\phi^{2}\right],
\end{equation}
whose Weyl tensor is zero because of conformal flatness.
According to the second section, the expression for the torsion scalar for the flat FLRW metric reduces to
\begin{equation}\label{eq30}
T=-6H^{2},
\end{equation}
 where $H=\frac{\dot{a}}{a}$ is the Hubble parameter. Note that the equation  (\ref{eq30}) holds only if a diagonal FRW tetrad is chosen, since different tetrads giving the same metric lead to different results in $f(T)$ gravity \cite{Boh}.  Apparently, the torsion scalar is thoroughly time-dependant, thus we are supposed to be concerned just about time derivatives of $T$ in $D_{\mu\nu}$.

The vector field normalization implies that $V_{\alpha}V^{\alpha}=\epsilon$ and also we have $E=-V_{\alpha}u^{\alpha}$, $\eta_{\alpha}u^{\alpha}=\eta_{\alpha}V^{\alpha}=0$, $\eta_{0}u^{0}=0$. Not only these mentioned conditions but also the non-vanishing components of $S$ tensor will be used to extract the final result for the action of operator $D_{\mu\nu}$ on $f_{T}$. Therefore, after cumbersome calculations we obtain
\begin{widetext}
\begin{eqnarray}\label{eq31}
\frac{1}{2f_{T}}\left((\delta^{\alpha}_{\gamma}D_{\delta\beta}-\delta^{\alpha}_{\delta}D_{\gamma\beta}+
g_{\beta\delta}D^{\alpha}_{\gamma}-g_{\beta\gamma}D^{\alpha}_{\delta})f_{T}\right)V^{\beta}V^{\delta}\eta^{\gamma}=
-24H^{2}\dot{H}f_{TT}(E^{2}+2\epsilon)\eta^{\alpha},
\end{eqnarray}
\end{widetext}
where have used $S^{1}_{10}=S^{2}_{20}=S^{3}_{30}=-2H(t)$. Thus, we can write the reduced expression for $R^{\alpha}\,_{\beta\gamma\delta}V^{\beta}\eta^{\gamma}V^{\delta}$
as follows
\begin{eqnarray}\label{eq32}
R^{\alpha}\,_{\beta\gamma\delta}V^{\beta}\eta^{\gamma}V^{\delta}&=&\frac{1}{2f_{T}}\Big[(\kappa(\rho+p)-24H^{2}\dot{H}f_{_{TT}})E^{2}+
\epsilon\Big(\frac{2\kappa\rho}{3}-\frac{f}{3}+\frac{Tf_T}{3}\Big)\Big]\eta^{\alpha},
\end{eqnarray}
which is the generalized {\it Pirani} equation. Now, we can write the GDE in $f(T)$ gravity model
\begin{eqnarray}\label{eq33}
\frac{D^{2}\eta^{\alpha}}{D\nu^{2}}&=&-\frac{1}{2f_{T}}\Big[(\kappa(\rho+p)-24H^{2}\dot{H}f_{_{TT}})E^{2}+
\epsilon\Big(\frac{2\kappa\rho}{3}-\frac{f}{3}+\frac{Tf_T}{3}\Big)\Big]\eta^{\alpha}.
\end{eqnarray}

Apparently, as a result of homogeneity and isotropy of FLRW metric, in this equation only the magnitude of the deviation vector $\eta^{\alpha}$ is changed along the geodesics. Whereas in anisotropic universe, like Bianchi $\textsc{I}$, we can also infer a change in the direction of the deviation vector, as described in \cite{cac}.
\subsection{Direct method with FLRW background }
Once we assume an FLRW metric (tetrad), then the Riemann tensor can be calculated. This
means that the geodesic deviation equation (\ref{eq12}) can be straightforwardly written in terms of $H$ and its derivative. We can
then use the $f(T)$ gravity cosmological equations (\ref{eq39}) and (\ref{eq40})  (see below) to connect $H$ and
$\dot H$ to the cosmological sources. Following this way, we can recover equation (\ref{eq32}) as follows. By taking radial spatial part, for suitable choice of local coordinates, we can write LHS of the equation  (\ref{eq32}) for $\alpha=r$ as follows
\begin{equation}\label{eq411}
R^{r}\,_{\beta\gamma\delta}V^{\beta}\eta^{\gamma}V^{\delta}=R^{r}\,_{t\gamma t}V^{t}\eta^{\gamma}V^{t}+R^{r}\,_{r\gamma r}V^{r}\eta^{\gamma}V^{r}+
R^{r}\,_{\theta \gamma  \theta}V^{\theta}\eta^{\gamma}V^{\theta}+R^{r}\,_{\phi \gamma  \phi}V^{\phi}\eta^{\gamma}V^{\phi}.
\end{equation}
By considering non-vanishing components of the Riemann tensor for the FLRW metric, we must take $\gamma=r$. Thus, the equation  (\ref{eq411}) can be written as
\begin{equation}\label{eq412}
R^{r}\,_{\beta\gamma\delta}V^{\beta}\eta^{\gamma}V^{\delta}=R^{r}\,_{trt}V^{t}V^{t}\eta^{r}+R^{r}\,_{rrr}~g^{rr}V_{r}V^{r}\eta^{r}+
R^{r}\,_{\theta r\theta}~g^{\theta\theta}V_{\theta}V^{\theta}\eta^{r}+
R^{r}\,_{\phi r\phi}~g^{\phi\phi}V_{\phi}V^{\phi}\eta^{r}=(-\dot{H}E^{2}+\epsilon H^{2}) \eta^{r},
\end{equation}
where we have used $V^{t}V^{t}=E^{2}$, $V_{i}\eta^{i}=0$, $R^{r}\,_{rrr}=0$, $R^{r}\,_{\theta r\theta}=r^{2}\dot{a}^{2}$, $R^{r}\,_{\phi r\phi}=\dot{a}^{2}r^{2}sin^{2}\theta$ and $R^{r}\,_{ttr}=\frac{\ddot{a}}{a}$. Note that similar equations are also obtained for $\alpha=\theta$ and $\alpha=\phi$.

According to the equations (\ref{eq39}) and (\ref{eq40}) (see below) we have
\begin{equation}\label{eqee}
H^{2}=\frac{1}{2f_T}\left(\frac{\kappa\rho}{3}-\frac{f}{6}\right),
\end{equation}
and
\begin{equation}\label{eqeee}
\dot{H}=-\frac{1}{2f_T}\left(\kappa(\rho+p)+4\dot{H}Tf_{TT}\right).
\end{equation}
By substituting the equation s (\ref{eqee}) and (\ref{eqeee}) into the equation  (\ref{eq412}) we can obtain the generalized {\it Pirani} equation (\ref{eq32})  by means of the direct approach as
\begin{eqnarray}\label{eqeeee}
R^{\alpha}\,_{\beta\gamma\delta}V^{\beta}\eta^{\gamma}V^{\delta}&=&\frac{1}{2f_{T}}\Big[(\kappa(\rho+p)-24H^{2}\dot{H}f_{_{TT}})E^{2}+
\epsilon\Big(\frac{2\kappa\rho}{3}-\frac{f}{3}+\frac{Tf_T}{3}\Big)\Big]\eta^{\alpha},
\nonumber
\end{eqnarray}
which results in the same geodesic deviation equation (\ref{eq33}). This means that we have found  the same results by two different approaches which
confirms the validity of the obtained geodesic deviation equation.

\subsection{ Fundamental observers with FLRW background}

Here, we are going to limit ourselves to the fundamental observers. In this particular case, we interpret $V^{\alpha}$ and $\nu$ (the affine parameter) as the four-velocity of the fluid $u^{\alpha}$  and $t$ (the proper time), respectively. Since we are treating with temporal geodesics we have $\epsilon=-1$ and also we fix the vector field normalization by $E=1$ which leads to
\begin{eqnarray}\label{eq34}
R^{\alpha}\,_{\beta\gamma\delta}u^{\beta}\eta^{\gamma}u^{\delta}=\frac{1}{2f_{T}}\Big[\frac{2\kappa\rho}{3}+\kappa p-24H^{2}\dot{H}f_{TT}+\frac{f}{6}\Big]\eta^{\alpha}.
\end{eqnarray}
We know that if $\eta_{\alpha}=\ell e_{\alpha}$, where $e_{\alpha}$ is parallel propagated along $t$, then the isotropy results in
\begin{eqnarray}\label{eq35}
\frac{De^{\alpha}}{Dt}=0,
\end{eqnarray}
from which we have
\begin{eqnarray}\label{eq36}
\frac{D^{2}\eta^{\alpha}}{Dt^{2}}=\frac{d^{2}\ell}{dt^{2}}e^{\alpha}.
\end{eqnarray}
By using (\ref{eq12}) and (\ref{eq34}) we can write
\begin{eqnarray}\label{eq37}
\frac{d^{2}\ell}{dt^{2}}=-\frac{1}{2f_{T}}\left[\frac{2\kappa\rho}{3}+\kappa p-24H^{2}\dot{H}f_{TT}+\frac{f}{6}\right]\ell,
\end{eqnarray}
which for the particular case $\ell=a(t)$ leads to
\begin{eqnarray}\label{eq38}
\frac{\ddot{a}}{a}=\frac{1}{f_{T}}\left[-\frac{\kappa\rho}{3}-\frac{\kappa p}{2}+12H^{2}\dot{H}f_{TT}-\frac{f}{12}\right].
\end{eqnarray}
This equation is nothing but a special case of the generalized Raychaudhuri equation. It is worth to mention here that the above generalized Raychaudhuri equation can be obtained by the standard forms of the modified  Friedmann equations in $f(T)$ gravity model for flat universe \cite{rafael2}

\begin{eqnarray}\label{eq39}
H^{2}=\frac{\kappa}{3}\left(\rho+\frac{1}{2\kappa}(2Tf_{T}-f-T)\right),
\end{eqnarray}

\begin{eqnarray}\label{eq40}
2\dot{H}+3H^{2}=-\kappa p-4\dot{H}Tf_{TT}-2\dot{H}f_{T}+2\dot{H}+Tf_{T}-\frac{f}{2}-\frac{T}{2}.
 \end{eqnarray}
The consistency between the modified  Friedmann equations in $f(T)$ gravity
for flat universe \cite{rafael2}
and the generalized Raychaudhuri equation for flat universe (\ref{eq38})
confirms the fact that the approach followed here is a correct one.\\

\subsection{ Null vector fields with FLRW background}

In this section,  we consider  the null past directed
vector fields,  namely  $V^{\alpha}=k^{\alpha}, k_{\alpha}k^{\alpha}=0$,
for which the equation (\ref{eq32}) reduces to

\begin{eqnarray}\label{eq41}
R^{\alpha}\,_{\beta\gamma\delta}k^{\beta}\eta^{\gamma}k^{\delta}=\frac{1}{2f_{T}}(\kappa(\rho+p)-24H^{2}\dot{H}f_{TT})E^{2}\eta^{\alpha}.
\end{eqnarray}
Actually, this is {\it Ricci focusing} in $f(T)$ gravity as is explained in
the following. By considering $\eta^{\alpha}=\eta e^{\alpha}$, $e_{\alpha}e^{\alpha}=1$, $\epsilon_{\alpha}u^{\alpha}=e_{\alpha}k^{\alpha}=0$ and also writing an aligned base parallel propagated $\frac{De^{\alpha}}{D\nu}=k^{\beta}\nabla_{\beta}e^{\alpha}=0$, we obtain the null GDE (\ref{eq33}) new form as follows

\begin{eqnarray}\label{eq42}
{\frac{d^{2}\eta}{d\nu^{2}}=-\frac{1}{2f_{T}}(\kappa(\rho+p)-24H^{2}\dot{H}f_{TT})E^{2}\eta.}
\end{eqnarray}
{According to the GR studied in \cite{2} all classes of past-directed null geodesics experience focusing if we have $\kappa(\rho+p)>0$. Therefore, in a particular case with the equation  of state $p=-\rho$ (cosmological constant) we can not recognize any focusing effect. Obviously (\ref{eq42}) shows the focusing condition for $f(T)$ gravity model
provided that }

\begin{eqnarray}\label{eq43}
\frac{\kappa(\rho+p)}{f_{T}}>\frac{24H^{2}\dot{H}f_{_{TT}}}{f_{T}}.
\end{eqnarray}

Now, we have an expression (\ref{eq42}) which can be written in terms of the redshift parameter $z$. To do this, we may write

\begin{eqnarray}\label{eq44}
\frac{d}{d\nu}=\frac{dz}{d\nu}\frac{d}{dz},
\end{eqnarray}

which results in

\begin{eqnarray}\label{eq45}
\frac{d^{2}}{d\nu^{2}}=\left(\frac{d\nu}{dz}\right)^{-2}\left[-\left(\frac{d\nu}{dz}\right)^{-1}\frac{d^{2}\nu}{dz^{2}}\frac{d}{dz}+\frac{d^{2}}{dz^{2}}\right].
\end{eqnarray}

Let us consider the null geodesics for which we have

\begin{eqnarray}\label{eq46}
(1+z)=\frac{a_{0}}{a}=\frac{E}{E_{0}}\rightarrow\frac{dz}{1+z}=-\frac{da}{a}.
\end{eqnarray}
Choosing $a_{0}=1$ (the current day value of the scale factor), leads to the following result for the past-directed case

\begin{eqnarray}\label{eq47}
dz=(1+z)\frac{1}{a}\frac{da}{d\nu}d\nu=(1+z)\frac{\dot{a}}{a}Ed\nu=E_{0}H(1+z)^{2}d\nu.
\end{eqnarray}
Thus we obtain
\begin{eqnarray}\label{eq48}
\frac{d\nu}{dz}=\frac{1}{E_{0}H(1+z)^{2}},
\end{eqnarray}
and so
\begin{eqnarray}\label{eq49}
\frac{d^{2}\nu}{dz^{2}}=-\frac{1}{E_{0}H(1+z)^{3}}\left[\frac{1}{H}(1+z)\frac{dH}{dz}+2\right],
\end{eqnarray}
where
\begin{eqnarray}\label{eq50}
\frac{dH}{dz}=\frac{d\nu}{dz}\frac{dt}{d\nu}\frac{dH}{dt}=-\frac{1}{H(1+z)}\frac{dH}{dt}.
\end{eqnarray}
where use has been made of
 $\frac{dt}{d\nu}=E=E_{0}(1+z)$. From the definition of Hubble parameter we can write
\begin{eqnarray}\label{eq51}
\dot{H}=\frac{\ddot{a}}{a}-H^{2}.
\end{eqnarray}
Using (\ref{eq38}), $\dot{H}$ becomes
\begin{eqnarray}\label{eq52}
\dot{H}=\frac{1}{f_{T}}\left[\frac{-\kappa \rho}{3}-\frac{\kappa p}{2}+12\dot{H}H^{2}f_{TT}-\frac{f}{12}\right]-H^{2},
\end{eqnarray}
thus
\begin{widetext}
\begin{eqnarray}\label{eq53}
\frac{d^{2}\nu}{dz^{2}}=-\frac{3}{E_{0}H(1+z)^{3}}\left[1+\frac{1}{3H^{2}f_{T}}\left(\frac{\kappa \rho}{3}+\frac{\kappa p}{2}-12\dot{H}H^{2}f_{TT}+\frac{f}{12}\right)\right].
\end{eqnarray}
\end{widetext}
Putting this result in (\ref{eq45}), we obtain
\begin{widetext}
\begin{eqnarray}\label{eq54}
\frac{d^{2}\eta}{d\nu^{2}}=E_{0}H(1+z)^{2}\left[\frac{d^{2}\eta}{dz^{2}}+\frac{3}{(1+z)}\left[1+\frac{1}{3H^{2}f_{T}}\left(\frac{\kappa \rho}{3}+\frac{\kappa p}{2}-12\dot{H}H^{2}f_{TT}+\frac{f}{12}\right)\right]\frac{d\eta}{dz}\right].
\end{eqnarray}
\end{widetext}
Finally, using (\ref{eq42}) the null GDE takes the following form
\begin{widetext}
\begin{eqnarray}\label{eq55}
\frac{d^{2}\eta}{dz^{2}}+\frac{3}{(1+z)}\left[1+\frac{1}{3H^{2}f_{T}}\left(\frac{\kappa \rho}{3}+\frac{\kappa p}{2}-12\dot{H}H^{2}f_{TT}+\frac{f}{12}\right)\right]\frac{d\eta}{dz}+\frac{\kappa(\rho+p)-24H^{2}\dot{H}f_{_{TT}}}{2H^{2}(1+z)^{2}f_{T}}\eta=0.
\end{eqnarray}
\end{widetext}
Matter and radiation contributions to $\rho$ and $p$ can be written respectively as

\begin{eqnarray}\label{eq56}
\kappa \rho=3H_{0}^{2}\Omega_{m0}(1+z)^{3}+3H_{0}^{2}\Omega_{r0}(1+z)^{4},\hspace{20mm}
\kappa p=H_{0}^{2}\Omega_{r0}(1+z)^{4},
\end{eqnarray}
where we have used $p_{m}=0$ and $p_{r}=\frac{1}{3}\rho_{r}$. By using the equation  (\ref{eq56}) the null GDE equation (\ref{eq55}) can be written as

\begin{eqnarray}\label{eq57}
\frac{d^{2}\eta}{dz^{2}}+P(H,\dot{H},z)\frac{d\eta}{dz}+Q(H,\dot{H},z)\eta=0,
\end{eqnarray}
where
\begin{widetext}
\begin{eqnarray}\label{eq58}
P(H,\dot{H},z)=\frac{\Omega_{m0}(1+z)^{3}+\frac{3}{2}\Omega_{r0}(1+z)^{4}+\frac{f}{12H_{0}^{2}}+(3f_{T}+
\frac{\dot{T}f_{TT}}{H})[\Omega_{m0}(1+z)^{3}+\Omega_{r0}(1+z)^{4}+\Omega_{DE}]}{f_{T}(1+z)[\Omega_{m0}(1+z)^{3}+\Omega_{r0}(1+z)^{4}+\Omega_{DE}]},
\end{eqnarray}
\end{widetext}
\begin{eqnarray}\label{eq59}
Q(H,\dot{H},z)=\frac{3\Omega_{m0}(1+z)^{3}+4\Omega_{r0}(1+z)^{4}+2\frac{\dot{T}f_{_{TT}}}{H}(\Omega_{m0}(1+z)^{3}+
\Omega_{r0}(1+z)^{4}+\Omega_{DE})}{2f_{T}(1+z)^{2}[\Omega_{m0}(1+z)^{3}+\Omega_{r0}(1+z)^{4}+\Omega_{DE}]},
\end{eqnarray}
in which  we have applied the following new form of (\ref{eq39})
\begin{eqnarray}\label{eq60}
H^{2}=H_{0}^{2}[\Omega_{m0}(1+z)^{3}+\Omega_{r0}(1+z)^{4}+\Omega_{DE}],
\end{eqnarray}
where $\Omega_{DE}$ has been defined as

\begin{eqnarray}\label{eq61}
\Omega_{DE}=\frac{1}{H_{0}^{2}}\left(\frac{Tf_{T}}{3}-\left(\frac{f+T}{6}\right)\right).
\end{eqnarray}
Note that in solving the equation  (\ref{eq57}), we must use \eqref{eq30}.
In order to check for the agreement of the above results with those of GR, we take the particular case $f(T)=T-2\Lambda$. As a result of this choice we have $f_{T}=1$ and $f_{TT}=0$.
Furthermore $\Omega_{DE}$ reduces to

\begin{eqnarray}\label{eq62}
\Omega_{DE}=\frac{1}{H_{0}^{2}}\left(\frac{T-2\Lambda}{3}-\left(\frac{T-2\Lambda+T}{6}\right)\right)=\frac{\Lambda}{3H_{0}^{2}}\equiv\Omega_{\Lambda}.
\end{eqnarray}
which can be written as the Friedmann equation in GR
\begin{eqnarray}\label{eq63}
H^{2}=H_{0}^{2}[\Omega_{m0}(1+z)^{3}+\Omega_{r0}(1+z)^{4}+\Omega_{\Lambda}].
\end{eqnarray}
  Hence we find the reduced expressions $P$ and $Q$ as follows

\begin{eqnarray}\label{eq64}
P(z)=\frac{\frac{7}{2}\Omega_{m0}(1+z)^{3}+4\Omega_{r0}(1+z)^{4}+2\Omega_{\Lambda}}{(1+z)[\Omega_{m0}(1+z)^{3}+\Omega_{r0}(1+z)^{4}+\Omega_{\Lambda}]},
\end{eqnarray}

\begin{eqnarray}\label{eq65}
Q(z)=\frac{3\Omega_{m0}(1+z)+4\Omega_{r0}(1+z)^{2}}{2[\Omega_{m0}(1+z)^{3}+\Omega_{r0}(1+z)^{4}+\Omega_{\Lambda}]}.
\end{eqnarray}Eventually, the GDE for null vector fields becomes
\begin{widetext}
\begin{eqnarray}\label{eq66}
\frac{d^{2}\eta}{dz^{2}}+\frac{\frac{7}{2}\Omega_{m0}(1+z)^{3}+4\Omega_{r0}(1+z)^{4}+
2\Omega_{\Lambda}}{(1+z)[\Omega_{m0}(1+z)^{3}+\Omega_{r0}(1+z)^{4}+\Omega_{\Lambda}]}\frac{d\eta}{dz}+
\frac{3\Omega_{m0}(1+z)+4\Omega_{r0}(1+z)^{2}}{2(\Omega_{m0}(1+z)^{3}+\Omega_{r0}(1+z)^{4}+\Omega_{\Lambda})}\eta=0.
\end{eqnarray}
\end{widetext}
In order to obtain the Mattig relation in GR \cite{sch}, we have to fix $\Omega_{\Lambda}=0$, $\Omega_{r0}+\Omega_{m0}=1$ which lead to

\begin{widetext}
\begin{eqnarray}\label{eq67}
\frac{d^{2}\eta}{dz^{2}}+\frac{\frac{7}{2}\Omega_{m0}(1+z)^{3}+4\Omega_{r0}(1+z)^{4}}{(1+z)[\Omega_{m0}(1+z)^{3}+
\Omega_{r0}(1+z)^{4}]}\frac{d\eta}{dz}+\frac{3\Omega_{m0}(1+z)+4\Omega_{r0}(1+z)^{2}}{2(\Omega_{m0}(1+z)^{3}+\Omega_{r0}(1+z)^{4})}\eta=0.
\end{eqnarray}
\end{widetext}
So, the equation (\ref{eq57}) gives us an opportunity to generalize the Mattig relation in $f(T)$ gravity.
 By considering the last result, we can infer the following expression for the observer area distance $r_{0}(z)$ \cite{sch}:

\begin{eqnarray}\label{eq68}
r_{0}(z)=\sqrt{\left|\frac{dA_{0}(z)}{d\Omega}\right|}=\left|\frac{\eta(z')|_{z}}{d\eta(z')/d\ell|_{z'=0}}\right|,
\end{eqnarray}
where $A_{0}$ is the area of the object and also $\Omega$ is the solid angle. Equipped with the $d/d\ell=E^{-1}_{0}(1+z)^{-1}d/d\nu=H(1+z)d/dz$ and setting
the deviation at $z=0$  to zero, clearly we have

\begin{eqnarray}\label{eq69}
r_{0}(z)=\left|\frac{\eta(z)}{H(0) d\eta(z')/dz'|_{z'=0}}\right|,
\end{eqnarray}
where $H(0)$ is the evaluated modified Friedmann equation (\ref{eq62}) at $z=0$.

\section{Numerical Solutions of GDE for  $f(T)$ gravity}

To solve null GDE (\ref{eq57}) in $f(T)$ gravity we should consider
$f(T)$ functional forms. One simple assumption  in simplifying the equation
(\ref{eq57})
 is $\dot T=0$. However, this reduces our model to Lambda-CDM since $\Omega_{DE}=Const$, as can be seen from the equation  (\ref{eq61}). To find new interesting features of $f(T)$
modified gravity we should consider models beyond the $\dot T=0$ assumption.
Actually, there are variety of  $f(T)$ functional forms  in the literature,
each of which having some interesting specific features. Even, some studies have been done to obtain viable $f(T)$ functional forms, using the Noether symmetry approach \cite{ata,Wei}.

Following the above discussion,  and for simplicity, first we  assume $\dot{T}=0$ which is equivalent to
  a constant $T=-6H^{2}=T_{0}$. Thus, the equation  (\ref{eq58}) can be rewritten as follows
\begin{eqnarray}\label{eq70}
 \frac{d^{2}\eta}{dz^{2}}+P(H,T_{0},z)\frac{d\eta}{dz}+Q(H,T_{0},z)\eta=0,
\end{eqnarray}
where
\begin{widetext}
\begin{eqnarray}\label{eq71}
P(H,T_{0},z)=\frac{\Omega_{m0}(1+z)^{3}+\frac{3}{2}\Omega_{r0}(1+z)^{4}+\frac{f(T_{0})}{12H_{0}^{2}}+3f_{T}(T_{0})[\Omega_{m0}(1+z)^{3}+
\Omega_{r0}(1+z)^{4}+\Omega_{DE}(T_{0})]}{f_{T}(T_{0})(1+z)[\Omega_{m0}(1+z)^{3}+\Omega_{r0}(1+z)^{4}+\Omega_{DE}(T_{0})]},
\end{eqnarray}
\end{widetext}
\begin{eqnarray}\label{eq72}
Q(H,T_{0},z)=\frac{3\Omega_{m0}(1+z)+4\Omega_{r0}(1+z)^{2}}{2f_{T}(T_{0})[\Omega_{m0}(1+z)^{3}+\Omega_{r0}(1+z)^{4}+\Omega_{DE}(T_{0})]},
\end{eqnarray}
and

\begin{eqnarray}\label{eq73}
\Omega_{DE}(T_{0})=\frac{1}{H_{0}^{2}}\left(\frac{T_{0}f_{T}(T_{0})}{3}-\left(\frac{f(T_{0})+T_{0}}{6}\right)\right).
\end{eqnarray}
To solve the equation  (\ref{eq70}), one can choose the  functional form of the $f(T_{0})$ as follows
\begin{eqnarray}\label{eq74}
f(T_{0})=\alpha T_0+\beta T_{0}^{2},
\end{eqnarray}
where $\alpha$ and $\beta$ are constants.  We can solve the equation  (\ref{eq70}) numerically which results in  $\eta(z)$ and $r_{0}(z)$ plotted in Fig.2,   as a function of $z$.
\begin{figure}[ht]
  \centering
  \includegraphics[width=3.5in]{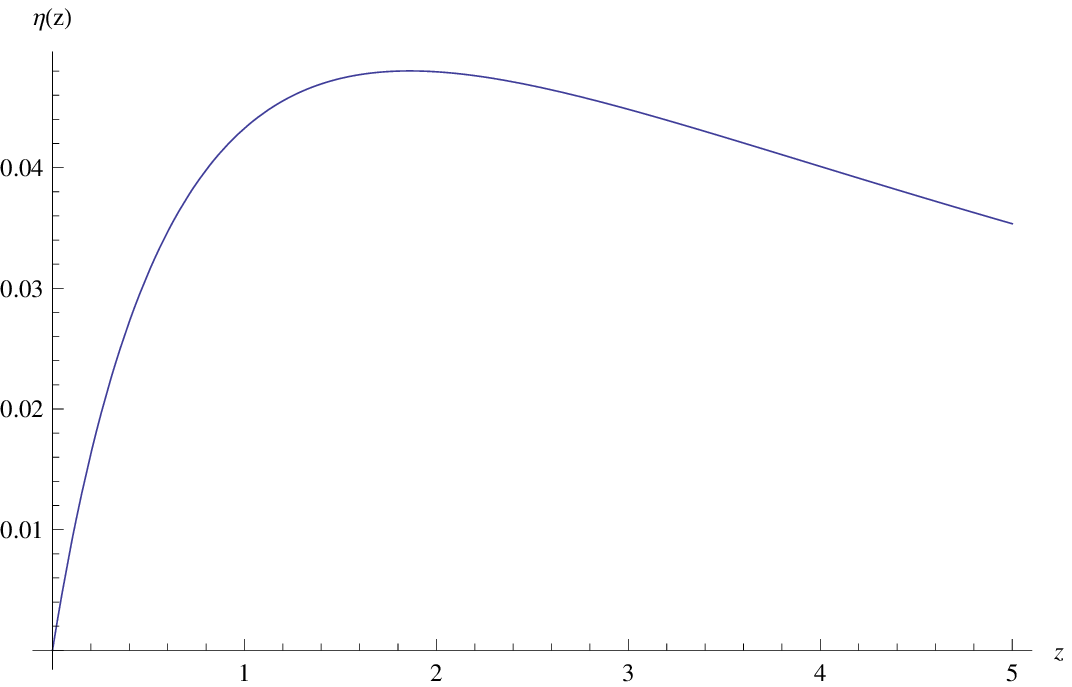}~~~~~
   \includegraphics[width=3.5in]{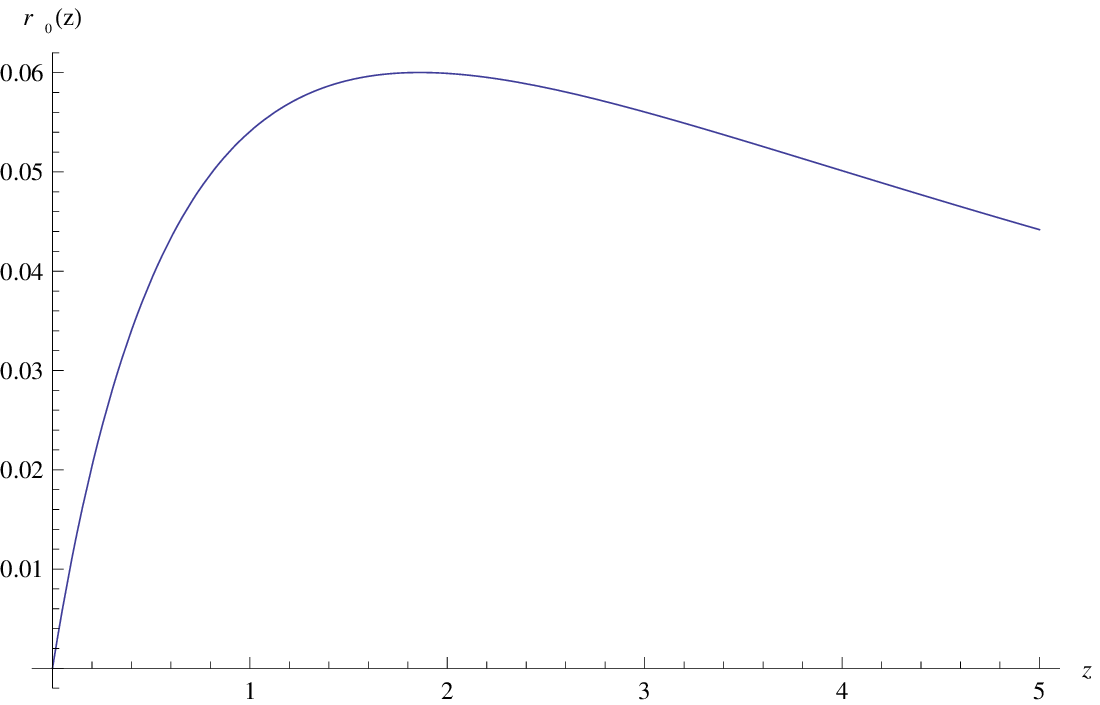}
  \caption{Plot of the deviation vector magnitude $\eta(z)$ (left) and plot of the observer area distance $r_0(z)$ (right). The parameter
values chosen are $H_0 = 80$ km/s/Mpc, $\Omega_{m0} = 0.3$, $\Omega_{r0} = 0$, $\Omega_{\Lambda} = 0.7$, and $d\eta(z)/dz|_{z=0}=0.1$.}
  \label{fig:1}
\end{figure}

\begin{figure}[ht]
  \centering
  \includegraphics[width=3.5in]{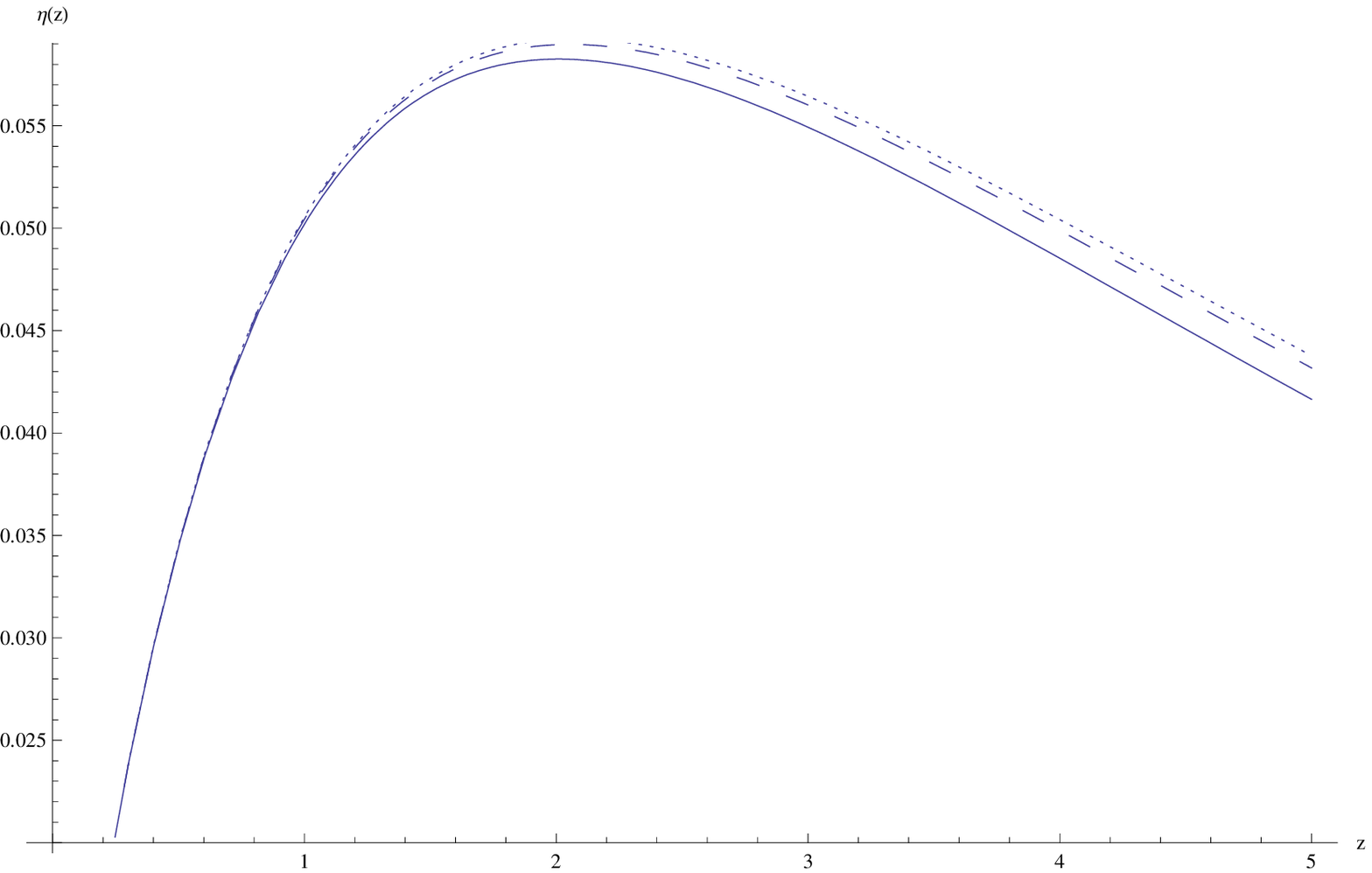}~~~~~
   \includegraphics[width=3.5in]{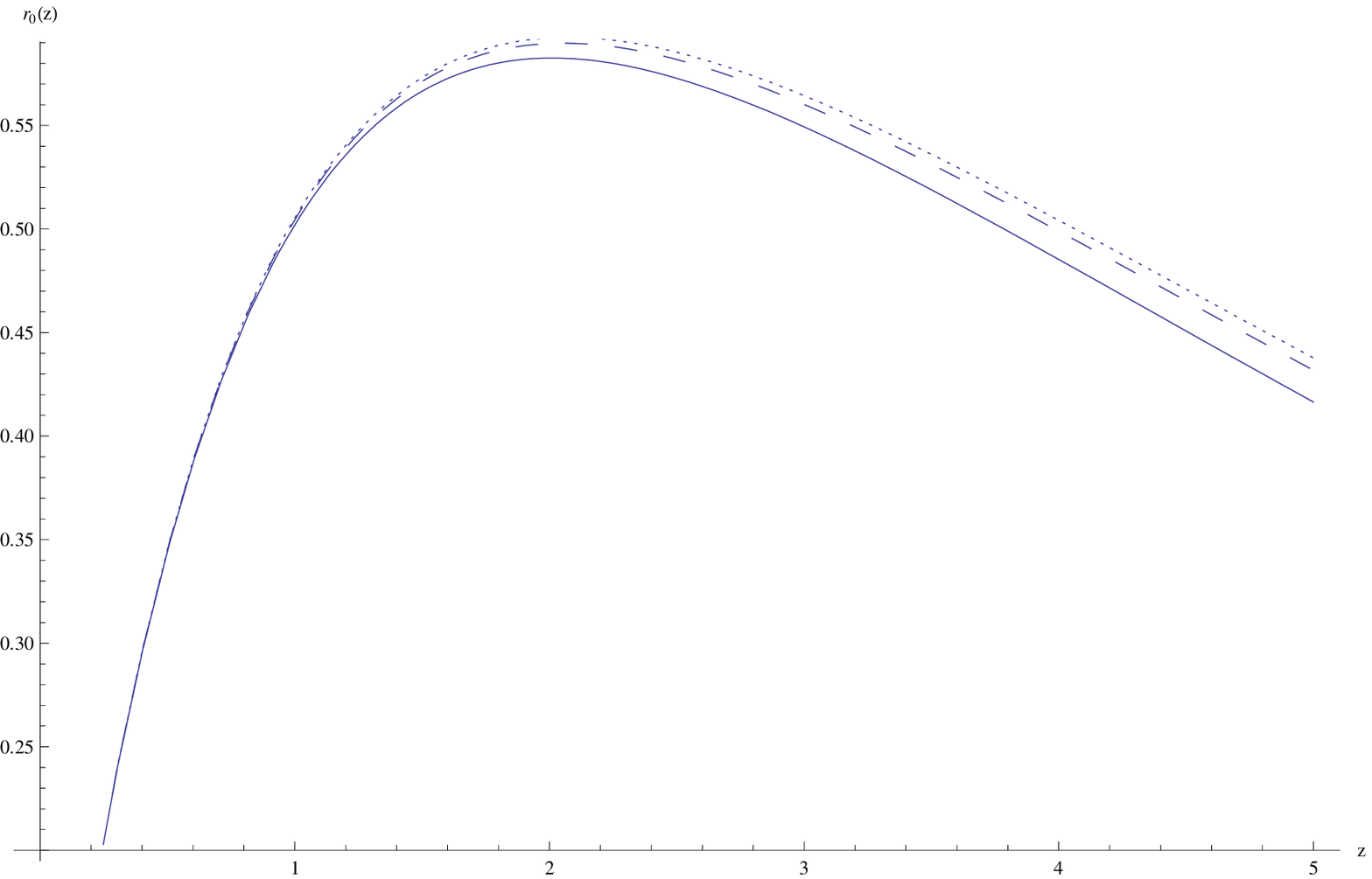}
  \caption{Plot of the deviation vector magnitude $\eta(z)$ (left) and plot of the observer area distance $r_0(z)$ (right). The parameter
values chosen are $H_0 = 80$ km/s/Mpc, $\Omega_{m0} = 0.3$, $a=0.9$, $b=0.2,
c=0.7$, $n=-1$ (solid), $n=-2$ (dotted), and $n=2$ (dashed), $\Omega_{r0} = 0$ and $d\eta(z)/dz|_{z=0}=0.1$.}
  \label{fig:2}
\end{figure}

\begin{figure}[ht]
  \centering
  \includegraphics[width=3.5in]{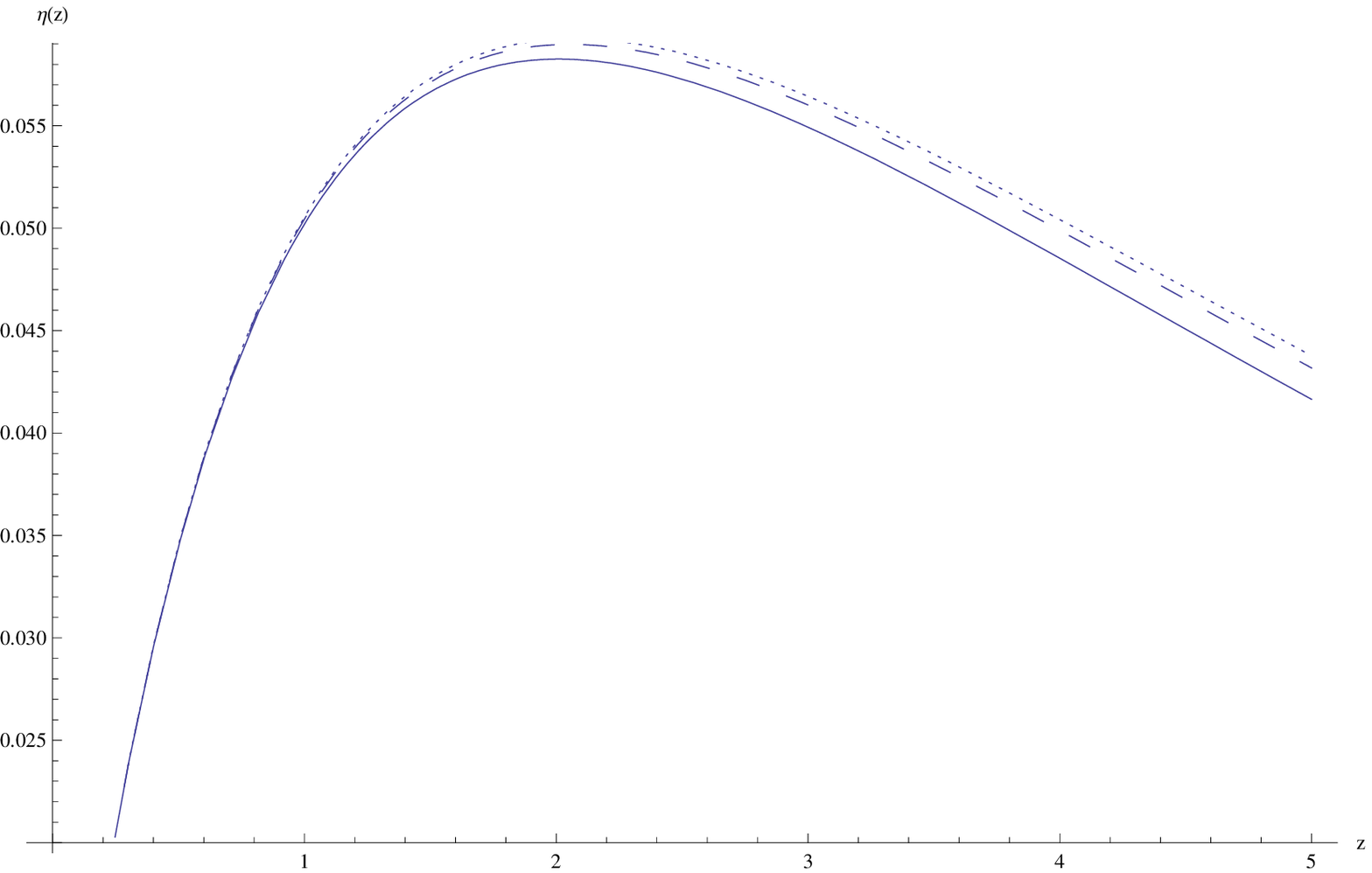}~~~~~
   \includegraphics[width=3.5in]{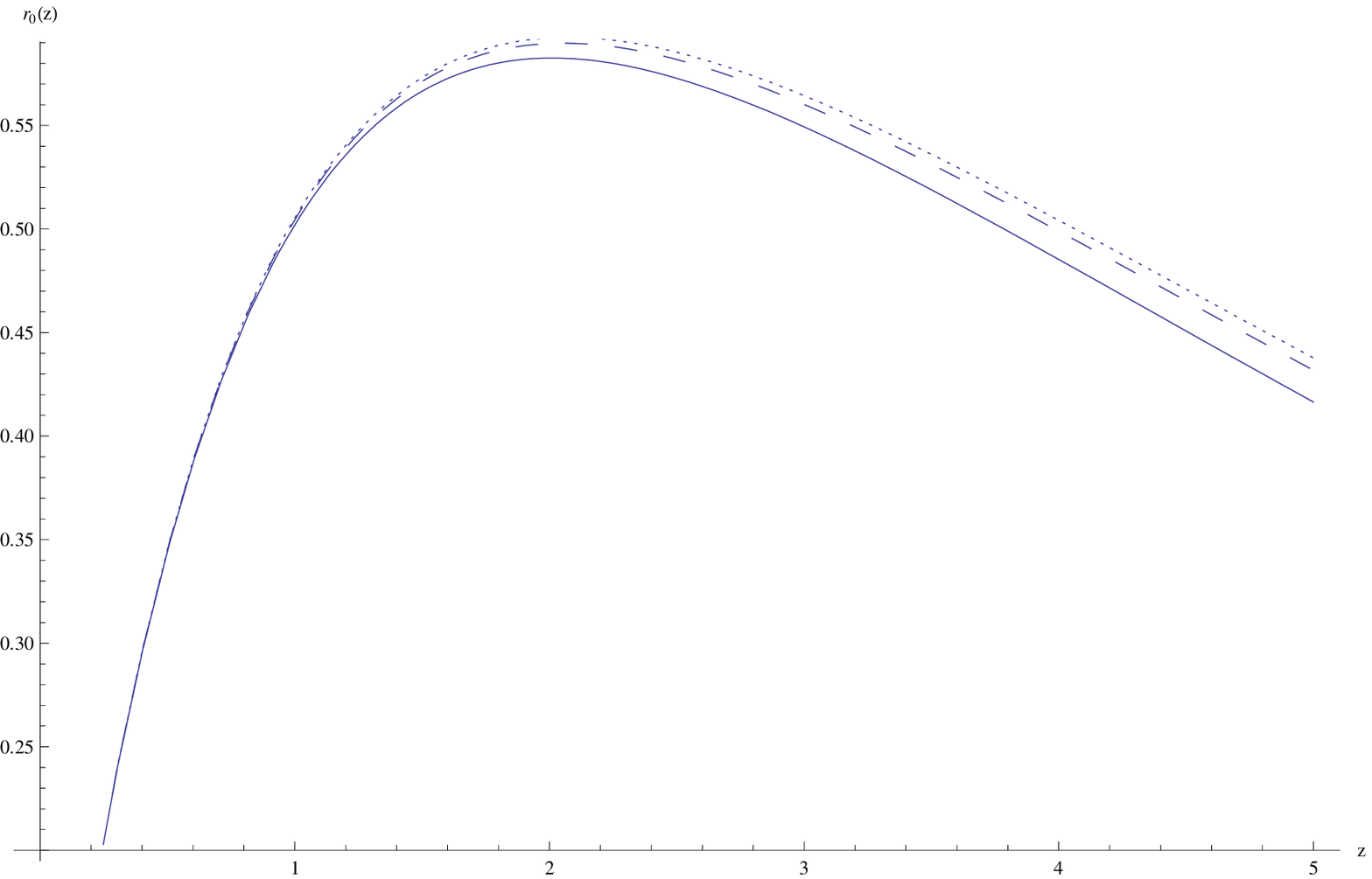}
  \caption{Plot of the deviation vector magnitude $\eta(z)$ (left) and plot of the observer area distance $r_0(z)$ (right). The parameter
values chosen are $H_0 = 80$ km/s/Mpc, $\Omega_{m0} = 0.3$, $a=0.9$, $b=0.2,
c=0.7$, $n=1$ (solid), $n=1.2$ (dotted), and $n=1.8$ (dashed), $\Omega_{r0} = 0$ and $d\eta(z)/dz|_{z=0}=0.1$.}
  \label{fig:3}
\end{figure}

Now, we assume $\dot{T}\neq0$. Since solving the GDE (\ref{eq57})   is not an easy task for any given $f(T)$ functional form, and   in order to much benefit of numerical calculations in solving  this complicate equation,
we just consider the following power law form as

\begin{eqnarray}\label{eq699}
f(T)=aT+bT^{n}+c~,
\end{eqnarray}
  which is well suited in most of the cosmological
implications of  $f(T)$ gravity \cite{Wei,ata},
or in a slightly modified form as
\begin{eqnarray}\label{eq699'}
f(T)=aT+\frac{bT^{n}}{1+T^{(n+1)}}+ c~,
\end{eqnarray}
where $a$, $b$, $c$ and $n$ are constants. One can rewrite the equation  (\ref{eq40}) as
\begin{eqnarray}\label{eq6699}
\frac{\dot{T}}{12H}=\frac{\kappa p+3H^{2}-Tf_{T}+\frac{f}{2}+\frac{T}{2}}{4Tf_{TT}+2f_{T}}.
\end{eqnarray}
Thus, by plugging the equation  (\ref{eq6699}) in the equations (\ref{eq58}) and (\ref{eq59}), we can solve the equation  (\ref{eq57}) numerically for suggested $f(T)$ forms (\ref{eq699}) and (\ref{eq699'}) which results in  $\eta(z)$ and $r_{0}(z)$ plotted, respectively in Fig.3 for $f(T)=aT+bT^{n}+c$ and in Fig.4 for $f(T)=aT+\frac{bT^{n}}{1+T^{(n+1)}}+ c$, as a function of $z$.

{For the model $T=-6H^{2}=T_{0}$, the behaviour of null geodesic deviation and observer area distance are the same
as those of $\Lambda$CDM model. This is expected because for $\dot T=0$ our
$f(T)$ model is reduced to $\Lambda$CDM model.
For the other suggested $f(T)$ models, namely $f(T)=aT+bT^{n}+c$, and $f(T)=aT+\frac{bT^{n}}{1+T^{(n+1)}}+ c$, the general behaviour of null geodesic deviation
and observer area distance are  similar to those of $\Lambda$CDM. At small values of $z<0.2$, the relative deviations and relative observer area distance with respect to the $\Lambda$CDM model are  almost ignorable. However,  for redshifts $z \gtrsim0.2$,
the geodesic deviation $\eta(z)$ and the area distance $r_{0}(z)$ corresponding
to the suggested $f(T)$ models are rather larger than those of $\Lambda$CDM model. This indicates that the suggested $f(T)$ models predict more acceleration
than $\Lambda$CDM model, for large values of redshift. }

\section{Conclusions}
In this paper, we have considered Geodesic Deviation Equation(GDE) in the GR equivalent of $f(T)$ gravity model. First, we have calculated the Ricci tensor and the Ricci scalar with the modified field equations in $f(T)$ gravity theory. Then, in FLRW universe, the Geodesic Deviation Equation corresponding to these GR equivalent quantities of $f(T)$ gravity is obtained. To show the consistency of our approach in constructing
the GR equivalent of $f(T)$ gravity, the generalized GDE and Pirani equations are recovered for  $f(T)=T-2\Lambda$.  We restricted our attention to extract
the GDE for two special cases, namely the fundamental observers and past directed null vector fields.
In these two cases we have found the Raychaudhuri equation, the generalized Mattig relation and the diametral angular distance differential  for $f(T)$ gravity theory. We have also obtained the  geodesic deviation $\eta(z)$ and the area distance $r_{0}(z)$ corresponding to two suggested $f(T)$ models and compared them with those of  $\Lambda$CDM model.
\section*{Acknowledgment}
We would like to thank the anonymous referee whose useful comments much improved the presentation of this manuscript. This work has been supported by a grant/research fund number
$217/D/10976$ from Azarbaijan Shahid Madani University.


\begin{thebibliography}{9}
\bibitem{1}C. W. Misner, K. S. Thorne, and J. H. Wheeler, {\it Gravitation}, W. H. Freeman and Company, 1973.
\bibitem{2} J. L. Synge, Ann. Math. {\bf35}, 705 (1934);\\
 F. A. E. Pirani, Acta Phys. Polon. {\bf15}, 389 (1956);\\ G. F. R. Ellis and H. Van Elst, [arXiv:gr-qc/9709060v1].
\bibitem{3} R. M. Wald, {\it General Relativity}, The University of Chicago Press, 1984;\\
 E. Poisson, {\it A Relativist's Toolkit - The Mathematics of Black-Hole Mechanics}, Cambridge University
Press, 2004.
\bibitem{6}A. Einstein, Math. Ann. {\bf102} 685 (1930);\\ A. Einstein,
Sitzungsber. Preuss. Akad. Wiss. Phys. Math. Kl. {\bf24} (1930)
401, physics/0503046;\\ C. Pellegrini and J. Pleba´nski, K. Dan. Vidensk.
Selsk. Mat. Fys. Skr. {\bf2} 2 (1962);\\ C. Møler, K.
Dan. Vidensk. Selsk. Mat. Fys. Skr. {\bf89} (1978) No. 13;\\
K. Hayashi and T. Nakano, Prog. Theor. Phys. {\bf38} (1967)
491;\\ K. Hayashi and T. Shirafuji, Phys. Rev. D
{\bf19} 3524 (1979); {\bf24} 3312 (1981).
\bibitem{Weitzenb}R. Weitzenb\"{o}ck, {\it Invarianten Theorie}, (Noordhoff, Groningen, 1923).

\bibitem{Ferraro:2006jd} R.~Ferraro and F.~Fiorini,
Phys.\ Rev.\ D \textbf{75}, 084031 (2007); 
R.~Ferraro, F.~Fiorini,
Phys.\ Rev.\ \textbf{D78}, 124019 (2008);

\bibitem{Ben09} G.~R.~Bengochea and R.~Ferraro,
Phys.\ Rev.\ D \textbf{79}, 124019 (2009).

\bibitem{Linder:2010py} E.~V.~Linder,
Phys.\ Rev.\ D \textbf{81}, 127301 (2010).

\bibitem{Myrzakulov:2010vz} S.~H.~Chen, J.~B.~Dent, S.~Dutta and
E.~N.~Saridakis, 
Phys.\ Rev.\ D \textbf{\ 83}, 023508 (2011);
K.~Bamba, C.~-Q.~Geng and C.~-C.~Lee,
arXiv:1008.4036 [astro-ph.CO]; 
R.~-J.~Yang, 
Europhys.\ Lett.\ \textbf{93}, 60001 (2011);
J.~B.~Dent, S.~Dutta, E.~N.~Saridakis,
JCAP \textbf{1101}, 009 (2011); 
Y.~Zhang, H.~Li, Y.~Gong, Z.~-H.~Zhu, 
JCAP \textbf{1107}, 015 (2011); 
Y.~-F.~Cai, S.~-H.~Chen, J.~B.~Dent, S.~Dutta, E.~N.~Saridakis,
Class.\ Quant.\ Grav.\ \textbf{28}, 215011 (2011);
M.~Sharif, S.~Rani, 
Mod.\ Phys.\ Lett.\ \textbf{A26}, 1657 (2011);
S.~Capozziello, V.~F.~Cardone, H.~Farajollahi and A.~Ravanpak,
Phys.\ Rev.\ D \textbf{84}, 043527 (2011);
K.~Bamba and C.~-Q.~Geng,
JCAP \textbf{1111}, 008 (2011); 
C.~-Q.~Geng, C.~-C.~Lee, E.~N.~Saridakis, Y.~-P.~Wu,
Phys.\ Lett.\ \textbf{B704}, 384 (2011);
H.~Wei, 
Phys.\ Lett.\ B \textbf{712}, 430 (2012); 
C.~-Q.~Geng, C.~-C.~Lee, E.~N.~Saridakis,
JCAP \textbf{1201}, 002 (2012); 
Y.~-P.~Wu and C.~-Q.~Geng,
Phys.\ Rev.\ D \textbf{86}, 104058 (2012); 
C.~G.~Bohmer, T.~Harko and F.~S.~N.~Lobo,
Phys.\ Rev.\ D \textbf{85}, 044033 (2012); 
H.~Farajollahi, A.~Ravanpak and P.~Wu,
Astrophys.\ Space Sci.\ \textbf{338}, 23 (2012);
M.~Jamil, D.~Momeni, N.~S.~Serikbayev and R.~Myrzakulov,
Astrophys.\ Space Sci.\ \textbf{339}, 37 (2012);
J.~Yang, Y.~-L.~Li, Y.~Zhong and Y.~Li,
arXiv:1202.0129 [hep-th]; 
K.~Karami and A.~Abdolmaleki,
JCAP \textbf{1204}, 007 (2012); 
C.~Xu, E.~N.~Saridakis and G.~Leon,
JCAP \textbf{1207}, 005 (2012); 
K.~Bamba, R.~Myrzakulov, S.~'i.~Nojiri and S.~D.~Odintsov,
arXiv:1202.4057 [physics.gen-ph]; 
H.~Dong, Y.~-b.~Wang and X.~-h.~Meng,
Eur.\ Phys.\ J.\ C \textbf{72}, 2002 (2012); 
K.~Bamba, S.~Capozziello, S.~Nojiri and S.~D.~Odintsov,
Astrophys.\ Space Sci.\ \textbf{342}, 155 (2012);
A.~Behboodi, S.~Akhshabi and K.~Nozari,
Phys.\ Lett.\ B \textbf{718}, 30 (2012); 
A.~Banijamali and B.~Fazlpour, 
Astrophys.\ Space Sci.\ \textbf{342}, 229 (2012);
D.~Liu and M.~J.~Reboucas, 
Phys.\ Rev.\ D \textbf{86}, 083515 (2012);
M.~E.~Rodrigues, M.~J.~S.~Houndjo, D.~Saez-Gomez and F.~Rahaman,
Phys.\ Rev.\ D \textbf{86}, 104059 (2012); 
Y.~-P.~Wu and C.~-Q.~Geng,
arXiv:1211.1778 [gr-qc]; S.~Chattopadhyay and A.~Pasqua,
Astrophys.\ Space Sci.\ \textbf{344}, 269 (2013);
M.~Jamil, D.~Momeni and R.~Myrzakulov,
Gen.\ Rel.\ Grav.\ \textbf{45}, 263 (2013);
K.~Bamba, J.~de Haro and S.~D.~Odintsov,
JCAP \textbf{1302}, 008 (2013); 
M.~Jamil, D.~Momeni and R.~Myrzakulov,
Eur.\ Phys.\ J.\ C \textbf{72}, 2267 (2012); 
J.~-T.~Li, C.~-C.~Lee and C.~-Q.~Geng,
Eur.\ Phys.\ J.\ C \textbf{73}, 2315 (2013); H.~M.~Sadjadi,
Phys.\ Rev.\ D \textbf{87}, 064028 (2013); 
A.~Aviles, A.~Bravetti, S.~Capozziello and O.~Luongo,
Phys.\ Rev.\ D \textbf{87}, 064025 (2013); 
Y.~C.~Ong, K.~Izumi, J.~M.~Nester and P.~Chen,
Phys.\ Rev.\ D \textbf{88}, 024019 (2013); 
K.~Bamba, S.~Nojiri and S.~D.~Odintsov,
arXiv:1304.6191 [gr-qc]; H.~Dong, J.~Wang and X.~Meng,
arXiv:1304.6587 [gr-qc]; 
G.~G.~L.~Nashed,
Astrophys.\ Space Sci.\ \textbf{348}, 591 (2013);
G.~Otalora,
JCAP \textbf{1307}, 044 (2013); 
J.~Amoros, J.~de Haro and S.~D.~Odintsov,
Phys.\ Rev.\ D \textbf{87}, 104037 (2013); 
G.~Otalora,
Phys.\ Rev.\ D \textbf{88}, 063505 (2013); 
C.~-Q.~Geng, J.~-A.~Gu and C.~-C.~Lee,
Phys.\ Rev.\ D \textbf{88}, 024030 (2013);
I.~G.~Salako, M.~E.~Rodrigues, A.~V.~Kpadonou, M.~J.~S.~Houndjo and
J.~Tossa, 
arXiv:1307.0730 [gr-qc]; 
A.~V.~Astashenok,
arXiv:1308.0581 [gr-qc]; 
M.~E.~Rodrigues, I.~G.~Salako, M.~J.~S.~Houndjo and J.~Tossa,
arXiv:1308.2962 [gr-qc]; K.~Bamba, S.~D.~Odintsov and D.~Saez-Gomez,
Phys.\ Rev.\ D \textbf{88}, 084042 (2013); 
K.~Bamba, S.~Capozziello, M.~De Laurentis, S.~Nojiri and D.~Saez-Gomez,
Phys.\ Lett.\ B \textbf{727}, 194 (2013); 
T.~Harko, F.~S.~N.~Lobo, G.~Otalora and E.~N.~Saridakis,
arXiv:1404.6212 [gr-qc];  
G.~Otalora, 
arXiv:1402.2256 [gr-qc]; K.~Bamba, S.~'i.~Nojiri and S.~D.~Odintsov,
arXiv:1401.7378 [gr-qc]; 
G.~Kofinas and E.~N.~Saridakis,
arXiv:1404.2249 [gr-qc];  

\bibitem{11} R. Ferraro, F. Fiorini, Phys. Lett. B\textbf{702}, 75 (2011).
\bibitem{111} P. Wu, H. W. Yu, Phys. Lett. B\textbf{693}, 415 (2010);
G. R. Bengochea, Phys. Lett. B\textbf{695}, 405 (2011).
\bibitem{1111} L. Iorio and E. N. Saridakis, Mon. Not. Roy. Astron. Soc.
427, 1555 (2012).

\bibitem{Wang} T. Wang, Phys. Rev. D\textbf{84}, 024042 (2011); R. -X. Miao,
M. Li and Y. -G. Miao, JCAP \textbf{1111}, 033 (2011); R. Ferraro,
F. Fiorini, Phys. Rev. D \textbf{84}, 083518 (2011).
\bibitem{Mussa} C. G. Boehmer, A. Mussa and N. Tamanini, Class.
Quant. Grav. \textbf{28}, 245020 (2011).
\bibitem{MH} M. H. Daouda, M. E. Rodrigues and M. J. S. Houndjo,
Eur. Phys. J. C \textbf{71}, 1817 (2011);\\ M. H. Daouda,M. E. Rodrigues and M. J. S. Houndjo, Eur. Phys. J.
C \textbf{72}, 1890 (2012).
\bibitem{sari} P. A. Gonzalez, E. N. Saridakis and Y. Vasquez, JHEP
\textbf{1207}, 053 (2012); S. Capozziello, P. A. Gonzalez,
E. N. Saridakis and Y. Vasquez, JHEP \textbf{1302}, 039 (2013);
G. G. L. Nashed, Gen. Rel. Grav. \textbf{45}, 1878 (2013);
K. Atazadeh and M. Mousavi, Eur. Phys. J. C \textbf{72}, 2272
(2012); G. G. L. Nashed, Phys. Rev. D \textbf{88}, 104034 (2013);
G. G. L. Nashed, Europhys. Lett. \textbf{105}, 10001 (2014).
\bibitem{Bas} S. Nesseris, S. Basilakos, E. N. Saridakis and
L. Perivolaropoulos, Phys. Rev. D \textbf{88}, 103010 (2013).

\bibitem{Wei} H. Wei, X. -J. Guo and L. -F. Wang, Phys. Lett. B \textbf{707},
298 (2012).
\bibitem{ata} K. Atazadeh and F. Darabi, Eur. Phys. J. C \textbf{72}, 2016
(2012).
\bibitem{Bas2} S. Basilakos, S. Capozziello, M. De Laurentis, A.
Paliathanasis and M. Tsamparlis, Phys. Rev. D. {\bf88}, 103526 (2013).
 \bibitem{capo2}S. Capozziello, R. De Ritis, C. Rubano and P. Scudellaro,
Riv. Nuovo Cim. \textbf{19N4}, 1 (1996); \\A. Paliathanasis, S. Basilakos, E.N. Saridakis, S. Capozziello,
 K. Atazadeh, F. Darabi, M. Tsamparlis, Phys. Rev. D {\bf89}, 104042 (2014).
\bibitem{fr1}A. Guarnizo, L. Castaneda, J. M. Tejeiro, Gen. Rel. Grav. {\bf43},  2713 (2011);\\
A. de la Cruz-Dombriz, P. K. S. Dunsby, V. C. Busti, S. Kandhai, Phys. Rev. D {\bf89},  064029 (2014), arXiv:1312.2022;\\
A. Guarnizo, L. Castaneda, J. M. Tejeiro, arXiv:1402.3196.
\bibitem{fr2}F. Shojai and A. Shojai, Phys. Rev. D {\bf78}, 104011 (2008).
\bibitem{TR-GR}R. Aldrovandi, and J. G. Pereira, {\it Teleparallel Gravity
- An Introduction} (Springer-Verlag 2013).

\bibitem{barrow}B. Li, T.P. Sotiriou and J. D. Barrow, Phys. Rev. D {\bf83}, 064035 (2011), [arXiv:1010.1041].
\bibitem{barrow2}B. Li, T. P. Sotiriou and J. D. Barrow, Phys. Rev. D {\bf83}, 104017 (2011), [arXiv:1103.2786].
\bibitem{barrow3}T. P. Sotiriou, B. Li and J. D. Barrow, Phys. Rev. D {\bf83},
104030 (2011), [arXiv:1012.4039].
\bibitem{reb}Di Liu and M. J. Reboucas, Phys. Rev. D {\bf86}, 083515 (2012), [arXiv:1207.1503].
\bibitem{linder} E.V. Linder, Phys. Rev. D {\bf81}, 127301 (2010), [arXiv:1005.3039].
\bibitem{rafael2}G.R. Bengochea and R. Ferraro, Phys. Rev. D {\bf79},  124019 (2009), [arXiv:0812.1205].
\bibitem{ellis}G. F. R. Ellis and H. Van Elst, [arXiv:gr-qc/9812046v5].
\bibitem{Boh}N.~Tamanini and C.~G.~Bohmer, 
Phys.\ Rev.\ D \textbf{86}, 044009 (2012).
\bibitem{cac}D. L. Caceres, L. Casta~neda, J. M. Tejeiro, J.
Phys. Conf. Ser. {\bf229},  012076 (2010), [arXiv:0912.4220v1].
\bibitem{sch}P. Schneider, J. Ehlers and E. E. Falco, {\it Gravitational Lenses}, (Springer-Verlag, 1999).
\end{thebibliography}
\end{document}